\documentclass[12pt]{article}
\usepackage{epsfig,psfrag}
\catcode`@=11

\def\@citex[#1]#2{\if@filesw\immediate\write\@auxout{\string\citation{#2}}\fi
  \def\@citea{}\@cite{\@for\@citeb:=#2\do
    {\@citea\def\@citea{,\penalty\@m}\@ifundefined
      {b@\@citeb}{{\bf ?}\@warning
       {Citation `\@citeb' on page \thepage \space undefined}}%
\hbox{\csname b@\@citeb\endcsname}}}{#1}}

\def\citer{\@ifnextchar [{\@tempswatrue\@citexr}{\@tempswafalse\@citexr[]}}

\def\@citexr[#1]#2{\if@filesw\immediate\write\@auxout{\string\citation{#2}}\fi
  \def\@citea{}\@cite{\@for\@citeb:=#2\do
    {\@citea\def\@citea{--\penalty\@m}\@ifundefined
       {b@\@citeb}{{\bf ?}\@warning
       {Citation `\@citeb' on page \thepage \space undefined}}%
\hbox{\csname b@\@citeb\endcsname}}}{#1}}

\begin{document}
 
\thispagestyle{empty}
\begin{flushright}
LMU 25/03\\
October 2003
\end{flushright}

\vspace*{1.5cm}
\centerline{\Large\bf Model Independent Bound on the Unitarity Triangle}
\vspace*{0.3cm}
\centerline{\Large\bf from CP Violation in $B \to\pi^+\pi^-$
 and $B \to\psi K_S$}
\vspace*{2cm}
\centerline{{\sc Gerhard Buchalla} and 
{\sc A. Salim Safir}}
\bigskip
\bigskip
\centerline{\sl Ludwig-Maximilians-Universit\"at M\"unchen, 
Sektion Physik,} 
\smallskip
\centerline{\sl Theresienstra\ss e 37, D-80333 Munich, Germany}

\vspace*{2.5cm}
\centerline{\bf Abstract}
\vspace*{0.3cm}

We derive model independent lower bounds on the CKM parameters
$(1-\bar\rho)$ and $\bar\eta$ as functions of the mixing-induced
CP asymmetry $S$ in $B\to\pi^+\pi^-$ and $\sin 2\beta$ from $B\to\psi K_S$.
The bounds do not depend on specific
results of theoretical calculations for the penguin contribution
to $B\to\pi^+\pi^-$. They require only the very conservative
condition that a hadronic phase, which vanishes in the heavy-quark
limit, does not exceed $90^\circ$ in magnitude. The bounds are 
effective if $-\sin 2\beta \leq S \leq 1$. Dynamical calculations
indicate that the limits on $\bar\rho$ and $\bar\eta$ are close
to their actual values.

\vspace*{0.4cm}


\vfill

\newpage
\pagenumbering{arabic}

In the standard model CP violation is described by a single complex
phase in the Cabibbo-Kobayashi-Maskawa (CKM) matrix of quark mixing.
After the first observation of CP violation with $K_L\to\pi^+\pi^-$
decays in 1964 \cite{CCFT}, the standard model scenario has passed
a crucial test with the recent discovery of a large CP asymmetry
in $B\to\psi K_S$ decays \cite{AA,HFAG} in agreement with
earlier predictions \cite{CSBS}.
Beyond this qualitative confirmation of standard model CP violation,
efforts are now increasingly directed towards precision studies, 
which could reveal the existence of new physics.
A large amount of experimental data continues to be collected
at current and future $B$-meson facilities, both at dedicated
$e^+e^-$ colliders and at hadron machines.

To be useful for precision tests of weak interactions the observables
must not depend on poorly understood hadronic quantities. Among the
rather few observables that are essentially free of such theoretical
uncertainties is the time-dependent CP asymmetry in $B\to \psi K_S$,
which measures $\sin 2\beta$ (where $\beta$ is one of the angles of
the CKM unitarity triangle). Of comparable interest is the corresponding
CP asymmetry in $B\to\pi^+\pi^-$ decays.
However, a well-known problem here are penguin contributions in the
decay amplitude, which introduce hadronic physics into the CP
asymmetry and spoil a straightforward interpretation in terms of
weak phases. A method to eliminate hadronic input based on an isospin
analysis \cite{GL} appears to be very difficult for a practical
implementation \cite{HQ,CKMF}.

In this Letter we show how theoretically clean information on
CKM parameters can be obtained from the mixing-induced CP asymmetries in
$B\to\psi K_S$ and $B\to\pi^+\pi^-$, in spite of 
the penguin contributions in the second case. We derive lower bounds on
the unitarity triangle parameters $(1-\bar\rho)$ and $\bar\eta$
\cite{WBLO} that depend only on observables of CP violation in these
channels.

\vspace*{0.4cm}

The time-dependent CP asymmetry in $B\to\pi^+\pi^-$ decays
is defined by
\begin{eqnarray}\label{acppipi}
A^{\pi\pi}_{CP}(t) &=& 
\frac{B(B(t)\to\pi^+\pi^-)-B(\bar B(t)\to\pi^+\pi^-)}{
  B(B(t)\to\pi^+\pi^-)+B(\bar B(t)\to\pi^+\pi^-)} \nonumber \\
&=& - S\, \sin(\Delta m_B t) + C\, \cos(\Delta m_B t)
\end{eqnarray}
where
\begin{equation}\label{scxi}
S=\frac{2\, {\rm Im}\xi}{1+|\xi|^2} \qquad
C=\frac{1-|\xi|^2}{1+|\xi|^2} \qquad
\xi=e^{-2 i\beta}\,\frac{e^{-i\gamma}+P/T}{e^{+i\gamma}+P/T}
\end{equation}
In terms of the Wolfenstein parameters $\bar\rho$ and $\bar\eta$
the CKM phase factors read
\begin{equation}\label{gambetre}
e^{\pm i\gamma}=\frac{\bar\rho\pm i \bar\eta}{\sqrt{\bar\rho^2+\bar\eta^2}}
\qquad
e^{-2 i\beta}=\frac{(1-\bar\rho)^2 -\bar\eta^2 - 2 i\bar\eta(1-\bar\rho)}{
                     (1-\bar\rho)^2 + \bar\eta^2}
\end{equation}
The penguin-to-tree ratio $P/T$ can be written as
\begin{equation}\label{ptrphi}
\frac{P}{T}=\frac{r e^{i\phi}}{\sqrt{\bar\rho^2+\bar\eta^2}}
\end{equation}
The real parameters $r$ and $\phi$ defined in this way are
pure strong interaction quantities without further dependence
on CKM variables.

For any given values of $r$ and $\phi$ a measurement of $S$ 
defines a curve in the ($\bar\rho$, $\bar\eta$)-plane.
Using the relations above this constraint is given by the
equation
\begin{equation}\label{srhoeta}
S=\frac{2\bar\eta [\bar\rho^2+\bar\eta^2-r^2-\bar\rho(1-r^2)+
       (\bar\rho^2 +\bar\eta^2-1)r \cos\phi]}{((1-\bar\rho)^2+\bar\eta^2)
         (\bar\rho^2+\bar\eta^2+r^2 +2 r\bar\rho \cos\phi)}
\end{equation}

The current experimental results for $S$ and $C$ are
\begin{equation}
\begin{array}{cc}
S=+0.02\pm 0.34\pm 0.05 \quad ({\rm BaBar} \cite{BABAR1}) &
-1.23\pm 0.41 ^{+0.08}_{-0.07} \quad ({\rm Belle} \cite{BELLE1}) \\
 & \\
C=-0.30\pm 0.25\pm 0.04 \quad ({\rm BaBar} \cite{BABAR1}) &
-0.77\pm 0.27\pm 0.08 \quad ({\rm Belle} \cite{BELLE1}) 
\end{array}
\end{equation}
A recent preliminary update from BaBar gives \cite{HFAG}
\begin{equation}\label{scprel}
S=-0.40\pm 0.22\pm 0.03 \qquad C=-0.19\pm 0.19\pm 0.05
\end{equation}
The present results are still largely inconclusive, but
the accuracy should improve in the near future.
In the remainder of this article we will focus our
attention on $S$ and will not need $C$. We note, however, that
both quantities always fulfill the inequality $S^2+C^2\leq 1$.
The implications of $C$ will be considered elsewhere \cite{BS}.

The hadronic parameters $r$ and $\phi$ can be computed using
QCD factorization in the heavy-quark limit \cite{BBNS1,BBNS3}.
A recent analysis gives \cite{BS}
\begin{equation}\label{rphi}
r=0.107\pm 0.031 \qquad \phi=0.15\pm 0.25
\end{equation}
where the error includes an estimate of potentially important
power corrections. We quote these values in order to give an indication
of the typical size of $r$ and $\phi$. However, we will not make
use of the detailed prediction (\ref{rphi}).
Instead, we allow for an arbitrary value $r\geq 0$ and
shall only assume that the strong phase $\phi$ fulfills
\begin{equation}\label{phi90}
-\frac{\pi}{2}\leq\phi\leq\frac{\pi}{2}
\end{equation}
This is well justified since QCD factorization in the heavy-quark limit
implies that the strong interaction phase $\phi$ is systematically
suppressed, either by $\alpha_s$, for perturbative, hard rescattering,
or by $\Lambda_{QCD}/m_b$ for soft corrections.

After these preliminaries we now turn to the constraints on
the CKM parameters $\bar\rho$ and $\bar\eta$ that can be derived
from CP violation in $B\to\psi K_S$ and $B\to\pi^+\pi^-$ decays.
The angle $\beta$ of the unitarity triangle is given by
\begin{equation}\label{taus2b}
\tau\equiv\cot\beta=\frac{\sin 2 \beta}{1-\sqrt{1-\sin^2 2\beta}}
\end{equation}
where $\sin 2\beta$ is measured directly and with negligible
theoretical uncertainty \cite{HQ} from the mixing-induced 
CP asymmetry in $B\to\psi K_S$. 
A second solution with a positive sign in front of the
square root in (\ref{taus2b}) is excluded using other
constraints on the unitarity triangle, such as $|V_{ub}/V_{cb}|$
\cite{CKMF}.
The current world average \cite{HFAG} 
\begin{equation}\label{sin2bexp}
\sin 2\beta =0.739\pm 0.048
\end{equation}
implies
\begin{equation}\label{tauexp}
\tau=2.26\pm 0.22
\end{equation}
Given a value of $\tau$, $\bar\rho$ is related to $\bar\eta$
by
\begin{equation}\label{rhotaueta}
\bar\rho = 1-\tau\, \bar\eta
\end{equation}

The parameter $\bar\rho$ may thus be eliminated from $S$
in (\ref{srhoeta}), which can be solved for $\bar\eta$ to yield
\begin{eqnarray}\label{etb2}
\bar\eta &=& \frac{1}{(1+\tau^2)S} 
  \Bigg[(1+\tau S)(1+r \cos\phi)  \nonumber \\
 &&  \qquad -\sqrt{(1-S^2)(1+r\cos\phi)^2-
  S(1+\tau^2)(S+\sin 2\beta) r^2 \sin^2\phi}\Bigg]
\end{eqnarray}
If $S\geq 0$, the coefficient of $r^2\sin^2\phi$ under the square root is
negative, hence
\begin{equation}\label{etb3}
\bar\eta\geq\frac{1+\tau S-\sqrt{1-S^2}}{(1+\tau^2)S}(1+r\cos\phi)
\end{equation}
If $-\sin 2\beta\leq S\leq 0$, the same bound is obtained so that
(\ref{etb3}) holds for the entire range $-\sin 2\beta\leq S\leq 1$.
We note that this bound is still {\it exact} and requires no information
on the phase $\phi$. (The only condition is that $(1+r\cos\phi)$ is
positive, which is no restriction in practice.)

Assuming now (\ref{phi90}), we have $1+r\cos\phi \geq 1$ and
\begin{equation}\label{etabound}
\bar\eta\geq\frac{1+\tau S-\sqrt{1-S^2}}{(1+\tau^2)S}
\quad {\rm if}\quad -\sin 2\beta\leq S\leq 1
\end{equation}
This constraint is the main result of this paper.
We emphasize that the lower bound on $\bar\eta$ depends only on the
\begin{figure}[t]
\psfrag{S}{$S$}
\psfrag{etab}{$\bar\eta$}
\begin{center}
\psfig{figure=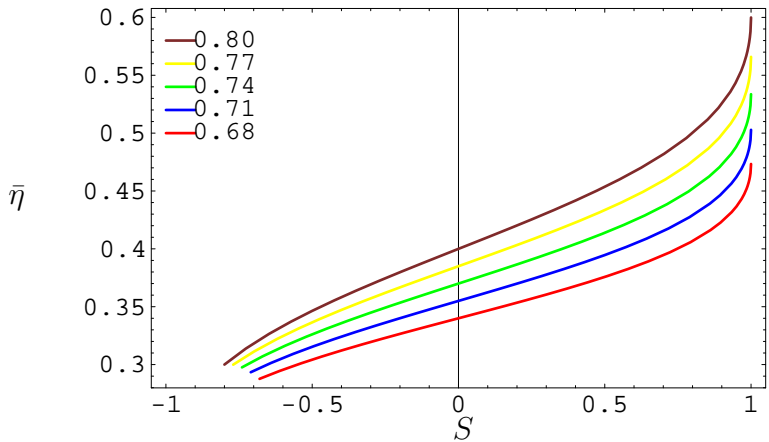}
\end{center}
{\it{\caption{Model-independent lower bound on $\bar\eta$ as a function
of $S$ for various values of $\sin 2\beta$
(increasing from bottom to top).  \label{fig:etabound}}}}
\end{figure}
observables $\tau$ and $S$ and is essentially free of hadronic
uncertainties. It holds in the standard model and it is effective under
the condition that $S$ will eventually  be measured in the interval
$[-\sin 2\beta,1]$. Since both $r$ and $\phi$ are expected to be
quite small, we anticipate that the lower limit (\ref{etabound})
is a fairly strong bound, close to the actual value of $\bar\eta$ itself.
We also note that the lower bound (\ref{etabound}) represents the
solution for the unitarity triangle in the limit of vanishing
penguin amplitude, $r=0$. In other words, the model-independent
bounds for $\bar\eta$ and $\bar\rho$ are simply obtained
by ignoring penguins and taking $S\equiv\sin 2\alpha$ when fixing
the unitarity triangle from $S$ and $\sin 2\beta$.

Let us briefly comment on the second solution for $\bar\eta$, which has 
the minus sign in front of the square root in (\ref{etb2}) replaced by a
plus sign. For positive $S$ this solution is always larger than
(\ref{etb2}) and the bound (\ref{etabound}) is unaffected.
For $-\sin 2\beta\leq S\leq 0$ the second solution gives a negative
$\bar\eta$, which is excluded by independent information on the
unitarity triangle (for instance from indirect CP violation in
neutral kaons ($\varepsilon_K$)).

\begin{figure}[t]
\psfrag{S}{$S$}
\psfrag{gamma}{$\gamma$ [deg]}
\begin{center}
\psfig{figure=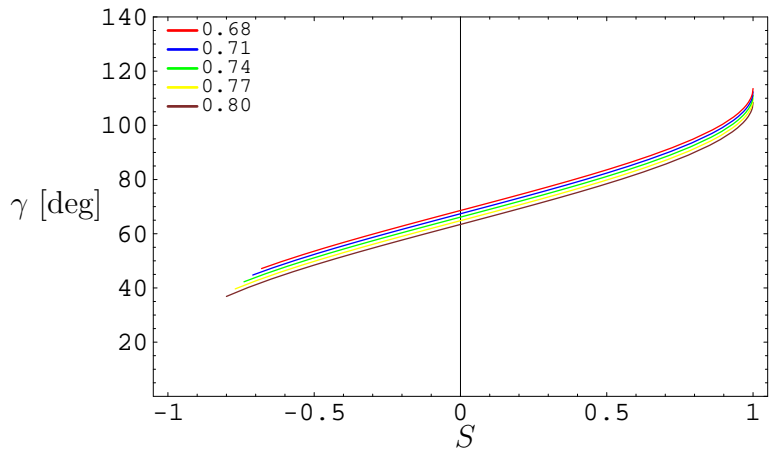}
\end{center}
{\it{\caption{Model-independent lower bound on $\gamma$ as a function
of $S$ for various values of $\sin 2\beta$
(decreasing from bottom to top).  \label{fig:gambound}}}}
\end{figure}

Because we have fixed the angle $\beta$, or $\tau$, the lower bound
on $\bar\eta$ is equivalent to an upper bound on $\bar\rho=1-\tau\bar\eta$.
The constraint (\ref{etabound}) may also be expressed as a lower bound
on the angle $\gamma$
\begin{equation}\label{gambound}
\gamma\geq \frac{\pi}{2}-
  \arctan\frac{S-\tau(1-\sqrt{1-S^2})}{\tau S+ 1-\sqrt{1-S^2}} 
\end{equation}
or a lower bound on $R_t$
\begin{equation}\label{rtbound}
R_t\equiv\sqrt{(1-\bar\rho)^2+\bar\eta^2}\geq
\frac{1+\tau S-\sqrt{1-S^2}}{\sqrt{1+\tau^2}S}
\end{equation}
In Figs. \ref{fig:etabound} and \ref{fig:gambound}
we represent, respectively, the lower bound on
$\bar\eta$ and $\gamma$ as a function of $S$ for various values of
$\sin 2\beta$. From Fig. \ref{fig:etabound} we observe that 
the lower bound on $\bar\eta$ becomes stronger as either
$S$ or  $\sin 2\beta$ increase. The sensitivity to $\sin 2\beta$
is less pronounced for the bound on $\gamma$.
Similarly to $\bar\eta$, the minimum allowed value for $\gamma$
increases with $S$. A lower limit $\gamma=90^\circ$ would be reached
for $S=\sin 2\beta$. 

\vspace*{0.4cm}

{\it To summarize}, we have demonstrated that theoretically clean
information can be extracted from CP violation in $B\to\pi^+\pi^-$
decays in the form of stringent limits on $\bar\rho$ and $\bar\eta$
even in the presence of penguin contributions. In order to make
an optimal use of these bounds, the quantities $\sin 2\beta$ and $S$
should be measured as precisely as possible. We hope that the analysis
suggested in this article will facilitate a transparent interpretation
and an efficient use of future data on CP violation in $B\to\pi^+\pi^-$.
Further discussions on this and related topics will be given in
a separate publication \cite{BS}.

\vspace*{0.5cm}

{\it Acknowledgement:} This work is supported in part by the Deutsche 
For\-schungs\-ge\-mein\-schaft (DFG) under contract BU 1391/1-2.

\vfill\eject

\end{document}